\documentstyle[11pt]{article}                       

\expandafter\ifx\csname amssym.def\endcsname\relax \else\endinput\fi
\expandafter\edef\csname amssym.def\endcsname{%
       \catcode`\noexpand\@=\the\catcode`\@\space}
\catcode`\@=11
\def\undefine#1{\let#1\undefined}
\def\newsymbol#1#2#3#4#5{\let\next@\relax
 \ifnum#2=\@ne\let\next@\msafam@\else
 \ifnum#2=\tw@\let\next@\msbfam@\fi\fi
 \mathchardef#1="#3\next@#4#5}
\def\mathhexbox@#1#2#3{\relax
 \ifmmode\mathpalette{}{\m@th\mathchar"#1#2#3}%
 \else\leavevmode\hbox{$\m@th\mathchar"#1#2#3$}\fi}
\def\hexnumber@#1{\ifcase#1 0\or 1\or 2\or 3\or 4\or 5\or 6\or 7\or 8\or
 9\or A\or B\or C\or D\or E\or F\fi}

\font\tenmsa=msam10
\font\sevenmsa=msam7
\font\fivemsa=msam5
\newfam\msafam
\textfont\msafam=\tenmsa
\scriptfont\msafam=\sevenmsa
\scriptscriptfont\msafam=\fivemsa
\edef\msafam@{\hexnumber@\msafam}
\mathchardef\dabar@"0\msafam@39
\def\dashrightarrow{\mathrel{\dabar@\dabar@\mathchar"0\msafam@4B}}
\def\dashleftarrow{\mathrel{\mathchar"0\msafam@4C\dabar@\dabar@}}

\def\ulcorner{\delimiter"4\msafam@70\msafam@70 }
\def\urcorner{\delimiter"5\msafam@71\msafam@71 }
\def\llcorner{\delimiter"4\msafam@78\msafam@78 }
\def\lrcorner{\delimiter"5\msafam@79\msafam@79 }
\def\yen{{\mathhexbox@\msafam@55}}
\def\checkmark{{\mathhexbox@\msafam@58}}
\def\circledR{{\mathhexbox@\msafam@72}}
\def\maltese{{\mathhexbox@\msafam@7A}}

\font\tenmsb=msbm10
\font\sevenmsb=msbm7
\font\fivemsb=msbm5
\newfam\msbfam
\textfont\msbfam=\tenmsb
\scriptfont\msbfam=\sevenmsb
\scriptscriptfont\msbfam=\fivemsb
\edef\msbfam@{\hexnumber@\msbfam}
\def\Bbb#1{{\fam\msbfam\relax#1}}
\def\widehat#1{\setbox\z@\hbox{$\m@th#1$}%
 \ifdim\wd\z@>\tw@ em\mathaccent"0\msbfam@5B{#1}%
 \else\mathaccent"0362{#1}\fi}
\def\widetilde#1{\setbox\z@\hbox{$\m@th#1$}%
 \ifdim\wd\z@>\tw@ em\mathaccent"0\msbfam@5D{#1}%
 \else\mathaccent"0365{#1}\fi}
\font\teneufm=eufm10
\font\seveneufm=eufm7
\font\fiveeufm=eufm5
\newfam\eufmfam
\textfont\eufmfam=\teneufm
\scriptfont\eufmfam=\seveneufm
\scriptscriptfont\eufmfam=\fiveeufm
\def\frak#1{{\fam\eufmfam\relax#1}}

\csname amssym.def\endcsname

\makeatletter
\def\section{\@startsection {section}{1}{\z@}{-3.5ex plus -1ex minus 
 -.2ex}{2.3ex plus .2ex}{\large\bf}}
\def\subsection{\@startsection{subsection}{2}{\z@}{-3.25ex plus -1ex minus 
 -.2ex}{1.5ex plus .2ex}{\normalsize\bf}}
\makeatother

\makeatletter
\@addtoreset{equation}{section}                        

\makeatother

\newcommand{\nc}{\newcommand}
\newcommand{\rnc}{\renewcommand}

\nc{\be}{\begin{equation}}
\nc{\ee}{\end{equation}}
\nc{\bea}{\begin{eqnarray}}
\nc{\eea}{\end{eqnarray}}

\nc{\trac}[2]{{\textstyle\frac{#1}{#2}}}

\nc{\ex}[1]{\mbox{e}^{\,\textstyle#1}}

\nc{\mat}[4]{\left(\begin{array}{cc}#1&#2\\#3&#4\end{array}\right)}

\nc{\som}[9]{\left(\begin{array}{ccc}#1&#2&#3\\#4&#5&#6\\#7&#8&#9%
\end{array}\right)}

\nc{\tr}{\mathop{\mbox{tr}}\nolimits}
\nc{\Tr}{\mathop{\mbox{Tr}}\nolimits}
\nc{\Det}{\mathop{\mbox{Det}}\nolimits}
\nc{\rk}{\mathop{\mbox{rk}}\nolimits}
\nc{\ad}{\mathop{\mbox{ad}}\nolimits}
\nc{\Ad}{\mathop{\mbox{Ad}}\nolimits}
\nc{\ra}{\rightarrow}
\nc{\Ra}{\Rightarrow}
\nc{\LRa}{\Leftrightarrow}
\nc{\ot}{\otimes}
\rnc{\ss}{\subset}
\nc{\noi}{\noindent}
\nc{\nul}{\noindent\underline}
\nc{\non}{\nonumber\\}
\nc{\RR}{{\Bbb R}}
\nc{\CC}{{\Bbb C}}
\nc{\ZZ}{{\Bbb Z}}

\rnc{\lg}{{\frak g}}
\nc{\lt}{{\frak t}}
\nc{\lk}{{\frak k}}
\nc{\lf}{{\frak f}}
\nc{\lh}{{\frak h}}
\nc{\bft}{{\frak t}}                          
\nc{\bfk}{{\frak k}}
\nc{\bfg}{{\frak g}}
\nc{\del}{\partial}
\nc{\dbar}{\bar{\del}}
\nc{\zb}{\bar{z}}
\nc{\az}{A_{z}}
\nc{\azb}{A_{\bar{z}}}
\nc{\bz}{B_{z}}
\nc{\bzb}{B_{\bar{z}}}
\nc{\g}{g^{-1}}
\nc{\dw}{\Delta_{W}}
\nc{\ddw}{\det\dw}
\nc{\Ddw}{\Det\dw}
\nc{\unA}{\underline{A}}                    
\nc{\unB}{\underline{B}}                    
\nc{\C}{{\cal A}/{\cal G}}                  
\nc{\A}[1]{{\cal A}^{#1}/{\cal G}^{#1}}     
\nc{\dx}{\dot{x}}
\rnc{\O}[2]{\Omega^{#1}({#2},\lg)} 
\nc{\wif}{Weyl integral formula}
\nc{\CS}{Chern-Simons}

\rnc{\d}{\delta}
\nc{\cb}{\bar{c}}
\nc{\f}{\phi}
\nc{\fb}{\bar{\phi}}
\nc{\Om}{\Omega} 
\nc{\p}{\psi}
\def\pb{\bar{\psi}}
\nc{\e}{\eta}
\nc{\eb}{\bar{\eta}}
\nc{\s}{\sigma}
\rnc{\o}{\omega}

\nc{\subs}[1]{{\vspace*{0.5cm}}%
{\noindent\underline{#1}}{\addcontentsline{toc}{subsection}{#1}}%
{\vspace*{0.3cm}}}

\newcommand{\ed}{\mbox{d}}
\newcommand{\maptor}{\Sigma_{\beta}}

\newcommand{\hsp}{\hspace{.25in},\hspace{.25in}}

\newcommand{\maps}{\rightarrow}

\newcommand{\Diff}[1]{\mbox{Diff}(#1)}

\rnc{\S}{\Sigma}
\nc{\Sa}{\S\times\{0\}}
\nc{\SI}{\S \times I}
\nc{\SS}{\S \times S^{1}}
\nc{\Sg}{\S_{g}}
\nc{\M}{{\cal M}}
\nc{\MF}{\M_{{\cal F}}}

\begin{document}
\global\parskip=4pt

\makeatletter
\begin{titlepage}
\begin{center}
{\LARGE\sc Solving Topological Field Theories on \\[4mm] 
Mapping Tori}\\
\vskip .3in
{\sc Matthias Blau}\footnote{e-mail: mblau@enslapp.ens-lyon.fr; supported 
by EC Human Capital and Mobility Grant ERB-CHB-GCT-93-0252}
\vskip .10in
Laboratoire de Physique Th\'eorique
{\sc enslapp}\footnote{URA 14-36 du CNRS,
associ\'ee \`a l'E.N.S. de Lyon,
et \`a l'Universit\'e de Savoie}\\
ENSLyon,
46 All\'ee d'Italie,\\
F-69364 Lyon CEDEX 07, France\\
\vskip .20in
{\sc Ian Jermyn}\footnote{e-mail: jermyn@ictp.trieste.it; supported 
by EC Human Capital and Mobility Grant ERB-CHB-ICT-93-0269}  and 
{\sc George Thompson}\footnote{e-mail: thompson@ictp.trieste.it}\\
\vskip .10in
ICTP\\
P.O. Box 586\\
34014 Trieste, Italy
\end{center}
\begin{abstract}
\noindent 
Using gauge theory and functional integral methods, we derive concrete
expressions for the partition functions of $BF$ theory and the $U(1|1)$
model of Rozansky and Saleur on $\SS$, both directly and
using equivalent two-dimensional theories.  We also derive the
partition function of a certain non-abelian generalization of the
$U(1|1)$ model on mapping tori and hence obtain explicit expressions for
the Ray-Singer torsion on these manifolds.  Extensions of these results
to $BF$ and Chern-Simons theories on mapping tori are also discussed.
The topological field theory actions of the equivalent two-dimensional
theories we find have the interesting property of depending explicitly
on the diffeomorphism defining the mapping torus while the quantum
field theory is sensitive only to its isomorphism class defining the
mapping torus as a smooth manifold.
\end{abstract}
\vfill
IC/95/341 \hfill hep-th/9605095 \hfill
{\small E}N{\large S}{\Large L}{\large A}P{\small P}-L-590/96
\end{titlepage}
\makeatother

\section{Introduction}

Topological field theories in three dimensions have been approached in
essentially three different ways. The first is through Witten's original
observation \cite{witt0} that one can get a handle on Chern-Simons 
theory by using
surgery and known results from conformal field theory (essentally from the
results of E. Verlinde \cite{ever}). There is also a combinatorial approach
as, for example, in defining the Turaev-Viro and Reshetikhin-Turaev invariants
\cite{TurVir}. This approach makes use of quantum groups and so is
intimately related to the first. The third method for computing is
perturbation theory \cite{AxSing, italians}. 

In this letter we will pursue a fourth approach, initiated in \cite{btver},
and examine and solve various topological field thories in three dimensions
using certain functional integral and gauge theoretic methods which (unlike
the usual perturbative approach) allow us to make maximal use of the
symmetries of the problem. We derive concrete, computable expressions 
for the Ray-Singer torsion on the manifolds and the partition functions of 
the theories considered, and not merely formal expressions. 
In particular, we will calculate the partition function of 
$BF$ theory on three-manifolds of the form $\SS$ by two different methods, 
analogous to those employed in \cite{btver} to solve Chern-Simons theory on 
$\SS$: directly in three dimensions (via Abelianisation) and 
using an equivalent two dimensional theory derived from $BF$ theory on
$\SI$ (the $BF$ analog of $G/G$ gauged Wess-Zumino-Witten models). 
As a further application of these techniques we also provide a simple
derivation of the partition function of the $U(1|1)$ model 
of Rozansky and Saleur \cite{RozSal} on $\SS$. 

In \cite{btver}, we had also
claimed that, in principle, these techniques are applicable to (topological)
gauge theories on mapping tori $\maptor$ associated to a diffeomorphism 
$\beta$ of $\S$. Here we illustrate this in the context of a certain 
natural non-Abelian generalization of the $U(1|1)$ model (a ``$G/T$'' model).
In particular, we obtain in this way topological field theories in two
dimensions which have the novel property of depending on (the equivalence
classes under isotopy and conjugation of) a diffeomorphism $\beta$. 
This thus provides us with a $\beta$-twisted
analog of the 2d/3d correspondence familiar from Chern-Simons theory.
A detailed investigation of these theories, however, 
as well as of those arising in a similar fashion from Chern-Simons and $BF$
theories will be left to a forthcoming publication \cite{us}. 

\noi The partition functions of the theories calculated here (integrals of
the Ray-Singer torsion over the moduli space of flat connections) are 
topological invariants of the manifolds considered. They may be thought of as
generalizations of the Johnson invariant to manifolds other than
homology three-spheres once we have dealt with the infinities that arise
along the way.

\section{The torus $\SS$ and the cylinder $\SI$}

Here we discuss $BF$ theory on $\SS$ and (as part of a second method of
solving the theory on $\SS$) on $\SI$, $I=[0,1]$. 
We also use these methods to give a
quick rederivation of the partition function of the $U(1|1)$ model of
Rozansky and Saleur \cite{RozSal} on $\SS$.

\subsection{$BF$ theory}

The general action for $BF$ theory (with a ``cosmological constant''
$\lambda$) on a three manifold $M$ is
\be
S(A,B) = \int_{M}B \wedge F(A) + \frac{\lambda}{3}B \wedge B \wedge B 
                                                    \label{eq:Action}
\ee
Here $A$ is a connection on a principal $G$-bundle over $M$, 
$B$ is a corresponding $L(G)$ (Lie algebra of $G$) valued 1-form and 
the integral is understood to include a trace. We note that for 
$\lambda > 0$ the theory can be transformed
into a Chern-Simons theory of $G\times G$, whereas for $\lambda < 0$ it is
the imaginary part of a Chern-Simons theory for $G_{\CC}$. 
Here we will discuss the $\lambda = 0$ case which can alternatively be
regarded as a Chern-Simons theory for a non-compact group commonly denoted
$IG$ (the tangent bundle group $TG$) \cite{witt2, witt3}.
The field equations tell us what the classical phase space is. The equation
from the $B$ variation
tells us immediately that we are dealing with the space of
flat connections on $M$. 
Fortunately the space of
solutions to the $A$ variation equation is equally easily characterized. 
This equation is just the linearized form of the A equation, telling us 
that we are dealing with the cotangent bundle to the space of flat 
connections as a
phase space. We have also to mod out by the gauge group but this preserves
the cotangent bundle structure and we are left with $T^{*}\M(M)$ as the
reduced phase space, where $\M(M)$ is the moduli space of flat connections
on $M$. General results \cite{general} tell us that we are
calculating the volume of $T^{*}\M(M)$ with the Ray-Singer torsion as the
measure.

We restrict the group in question to be compact, semi-simple and 
simply-connected, unless we specify $G=U(1)$. In either case we are dealing
only with trivial bundles, in the first case because that is all there is,
and in the second because of the flatness condition.
We also make use of an orthogonal decomposition of the Lie algebra:
\be
\lg = \lt \oplus \lk \label{decomp}
\ee

\noi where $\lt$ is the Cartan subalgebra. For manifolds $M$ of the form 
$\SS$ there is a particulary useful gauge choice using this decomposition, 
namely
\be
\dot{A}_{0}^{\lt}=0 \, , \; \; \; A_{0}^{\lk}=0 \label{gfA}
\ee
plus the condition that $A_{0}$ be compact. The reason that one cannot
fix to the gauge $A_{0}=0$ is that on the circle the holonomy is gauge
invariant and cannot be set to zero. The compactness condition arises as
follows. Even after we have imposed the conditions set out in
(\ref{gfA}) there are still `large' periodic gauge transformations
available which shift $A_{0}^{\lt}$ by elements of the integer lattice
$I$ of $\lt$. These have the form
\be
g(t) = g(0) \exp{t \gamma} \, , \; \; \; g(1) = g(0) \leftrightarrow
\gamma \in I \, , 
\ee
and shift $A_{0}^{\lt}$ by
\be
g^{-1}(t)A_{0}^{\lt}g(t) + g(t)^{-1}\partial_{0}g(t) = 
A_{0}^{\lt} + \gamma \, .
\ee
In these formula the group element $g(t) \in T $ may have $\S$
dependence ($T$ is the maximal torus of $G$ corresponding to the
decomposition (\ref{decomp})). We note that the conjugation of $A_{0}$ 
into the torus can
in general not be achieved globally and enforcing it introduces a sum over
all torus bundles, as described in \cite{btmap}. 

Using the above gauge conditions we can calculate an expression for the
partition function for $BF$ theory on $\SS$. 

\subsubsection{Solution}

We gauge fix $B$ by imposing $D_{0}B_{0}=0$. On the $\lt$ and $\lk$
components this condition becomes
\be
\partial_{0}B_{0}^{\lt} = 0 \hsp D_{0}B_{0}^{\lk} = 0\, .
\ee
We append to the action the ghost and gauge fixing terms
\be
\int_{\SS}\Lambda D_{0}B_{0} + \bar{c}D_{0}c + \bar{\rho}D_{0}^{2}\rho\, .
\ee
These give us determinants and delta functions
\be
\d(\partial_{0}B_{0}^{\lt}) \d(B_{0}^{\lk}) 
\Det{}_{\lk}(D_{0})\left.\right|_{0}^{-1}
\Det{}'_{\lt}(\partial_{0})\left.\right|_{0}
\Det{}_{\lk}(D_{0})\left.\right|_{0} \Det(D_{0}^{2})\left.\right|_{0}
\ee
where the $\left.\right|_{0}$ indicates evaluation on zero-forms. The prime
on the $\Det$s indicates that the zero modes are omitted in the evaluation.
We integrate out the non-constant $\unB$ modes to be left with an action
\be
\int_{\SS} B_{0}(\underline{\ed}\underline{a} 
			   + \frac{1}{2}[\underline{a},\underline{a}]) + \underline{b}
(\underline{\ed} + \ad (\underline{a}))a_{0}
\ee
plus the gauge fixing and ghosts and a new determinant 
$\Det'(D_{0})\left.\right|_{1}^{-1}$ where the $1$ indicates it is to be
evaluated on the space of one-forms on $\S$. The lower case letters in the
above equation indicate constant modes of the fields considered. The
$\underline{b}$ integral now contributes delta functions
\be
\d (\underline{\ed}a_{0})\d ([a_{0},\underline{a}^{\lk}])
            = \d (\underline{\ed}a_{0})\d (\underline{a}^{\lk})
              \Det{}_{\lk}(\ad(a_{0}))^{-1}\, .
\ee
Using all the delta functions at our disposal 
the action then reduces to 
\be
\int_{\SS} b_{0}^{\lt}\underline{\ed}\underline{a}^{\lt}\, .
\ee
This tells us that $b_{0}^{\lt}$ is constant and integral and thus the
$b_{0}^{\lt}$ integral contributes an overall factor of $\zeta(0)$ to 
the partition function.
Collecting the determinants from the ghosts and gauge fixing term and those
from above leaves us with an integral over 
$a_{0}$ (constant in space and time):
\be
\int da_{0} 
\frac{\Det{}_{\lk}(\partial_{0} + a_{0})\left.\right|_{0}^{2}}
	   {\Det{}_{\lk}(\partial_{0} + a_{0})\left.\right|_{1}}
\ee
which gives, after the usual regularization described in \cite{btver},
\be
Z = \int_{T} \det{}_{\lk}(1 - \Ad(t))^{\chi} \label{resultBFSS}
\ee

\noi where $t$ is an element in the maximal torus $T$ and $\chi$ is the
Euler character of $\S$ and we define $\Ad(g)\phi = g^{-1}\phi g$. Thus we 
obtain the Ray-Singer torsion on $S^{1}$ raised to the power of
the Euler character, i.e.\ the Ray-Singer torsion of $\SS$. 
Recall that the Chern-Simons partition function is 
the square root
of this result and note that the integral here is not cut off by the level 
as it is in the Chern-Simons case. We have been assuming no $\unB$ $D_{0}$ 
zero modes. If they do 
exist the determinants must be taken in the orthocomplement to the kernel of
the operators involved. We further note that we have discarded an infinity
that arises from the integral over $\underline{b}^{\lt}$. The harmonic modes
did not enter into the delta function on $\underline{\ed}a_{0}$, and thus we
are left with an integral of the form $\prod_{i=1}^{2g}\int\, db^{i}$ which
plainly contributes an infinity. We compare this with a formal result of
Witten \cite{witt1}, that the partition function should take the form
\be
Z = \int_{T_{A}{\cal M}}\, DB \int_{\cal M}\, T^{RS}_{M}(A)
\ee
where $T^{RS}_{M}(A)$ denotes the Ray-Singer torsion of $M$ 
with respect to the
flat connection $A$. The infinity that we find is that coming
from the integral over the tangent space, whereas our finite results
correspond to the integral over the moduli space of flat connections 
${\cal M}$. For homology three-spheres this is the Johnson invariant and, 
as mentioned in the introduction, it is a topological invariant of the 
manifold. This is the justification for our remark that the quantities we
are calculating are generalizations of this invariant to manifolds other
than homology three-spheres. 

\subsection{$BF$ theory on $\SI$}

As in \cite{btver} we will use the manifold $\SI$ to construct a
two-dimensional theory equivalent to $BF$ theory on $\SS$ by imposing
boundary conditions on $\SI$ and then
reconstructing the partition function on $\SS$ by integrating over these
boundary values while imposing a periodicity condition. The boundary values
are interpreted as fields in two dimensions. We also gauge fix the $\SS$
fields $A_{0}$ and $B_{0}$ by $A_{0}(x,t) = A_{0}(x)$ and 
$B_{0}(x,t) = B_{0}(x)$.  Our choice of boundary conditions is 
\be
\unA_{q}(0) = \unA  \hsp \unB_{q}(1) = \unB
\ee
\noi where the subscript $q$ now differentiates the dynamical fields from
the boundary conditions. This necessitates adding boundary terms to the 
action to have a
well-defined variational principle. It is easily seen that this term has the
form
\be
+\int_{\S \times \{1\}}\unB_{q} \wedge \unA_{q}\, .
\ee
Thus we are evaluating the following partition function:
\bea
Z_{\SS} &=& \int DA_{0}\, DB_{0}\, D\unA\, D\unB\, \delta(\partial_{0}A_{0})
	\delta(\partial_{0}B_{0}) \Det'(D_{0})\left.\right|_{0}^{2}\nonumber \\
    & & \times e^{-i\int_{\S}\unB\unA} Z_{\SI}[\unA, \unB, A_{0}, B_{0}]
	\label{BFpart}
\eea
where $Z_{\SI}$ is the partition function on $\SI$ with the given boundary
conditions and with $A_{0}$ and $B_{0}$ as background fields. (For
notational purposes we will initially denote these fields in the 
$\SI$ partition function by $A_{q\, 0} = A_{0}$ and $B_{q\, 0} = B_{0}$.)
One gets to this formula by thinking of the path integral on $\SS$ as
\be
\int D\unA \langle \unA, t\!=\!1 \mid \unA , t\!=\!0 \rangle = 
\int D\unA D\unB \langle \unA, t\!=\!1 \mid \unB
,  t\!=\!1\rangle \langle \unB, t\!=\!1 \mid \unA , t\!=\!0 \rangle\, ,
\ee
together with 
\be
\langle \unA, t\!=\!1 \mid \unB , t\!=\!1 \rangle = 
e^{-i \int_{\S}\unB\unA}\, .
\ee

\subsubsection{Reduction to a 2-dimensional theory}

We will evaluate the model on $\SI$. The action in this case is
\be
\int_{\SI} B_{q}F_{A_{q}} + \int_{\S \times [1]} \unB_{q}\unA_{q} \, .
\ee

Now write the action as
\be
\int_{\SI} B_{q}^{g}F_{A_{q}^{g}} + \int_{\S \times [1]} \unB_{q}\unA_{q}
\ee
and choose $g$ to solve $A^{g}_{q \, 0}=0$ with $g(1)=1$. Now send
$\unB_{q}
\rightarrow \unB_{q}^{g^{-1}}$ and $\unA_{q} \rightarrow \unA_{q}^{g^{-1}}$.
This has the effect of turning off $A_{q \, 0}$ in the action, which becomes
\be
\int_{\SI} B_{q}F_{(\unA_{q},0)}  + \int_{\S \times  [1]} \unB_{q}\unA_{q}
\, .
\ee
where now $B_{q\, 0} = g^{-1}B_{0}g$.
Of course we see the change of variables in the boundary data which now
reads
\be
\d(\unA_{q}(0)-\unA^{g})\d(\unB_{q}(1)-\unB)\, .
\ee
The aim now is to trivialise the dependence on $B_{q\, 0}$. We rewrite the
action as
\be
\int_{\SI}( B_{q}+d_{(\unA_{q},0)}\Lambda)F_{(\unA_{q},0)}
  + \int_{\S \times [1]} \unB_{q}\unA_{q} -\int_{\S \times [1]} \Lambda
F_{\unA_{q}} + \int_{\S \times [0]} \Lambda
F_{\unA_{q}} \, .
\ee
One picks $\Lambda$ to solve $B_{q\, 0} + \partial_{0} \Lambda =0$ with
$\Lambda(1) =0$. That is, one sets $\Lambda = \int_{t}^{1}B_{q\, 0}$.
We now change
variables, according to $\unB_{q} \rightarrow \unB_{q}
-\underline{\ed}_{\unA_{q}}\Lambda$. The net effect in the action is to set
$B_{q\, 0}$ to zero up to a boundary term,
\be
\int_{\S \times [0]} g^{-1}\f g F_{\unA_{q}} + \int_{\SI}
\unB_{q} \partial_{0} \unA_{q} + \int_{\S \times [1]}
\unB \unA_{q}\, .
\ee
where 
\be
\f = \int_{0}^{1}B_{q\, 0} = \frac{1}{\ad (A_{0})}(\Ad(g^{-1})-1)B_{0}.
\label{redefB}
\ee
The boundary conditions on $A_{q}$ and $B_{q}$ are not changed. The
delta function constraints on $\unA_{q}$ imply that it is equal to
$\unA^{g}$ while those on $\unB_{q}$ imply that this is $\unB$.

We have thus established that
\bea
Z[\unA,\unB,A_{0},B_{0}]  &=&  
e^{i \int_{\S \times [0]} g^{-1}\f g F_{\unA^{g}}} \times\non 
&& \int_{\unA^{g}}\, D\unA_{q} 
\int^{\unB} \, D\unB_{q} e^{i \int_{\SI} \unB_{q}\partial_{0} 
\unA_{q} +i \int_{\S \times [1]} \unB\unA_{q}} \, .
\eea
Notice that the path integral
\be
\int_{\unB:\unA^{g}}
D\unA_{q} D\unB_{q} e^{i \int_{\SI} \unB_{q}\partial_{0} 
\unA_{q} +i \int_{\S \times [1]} \unB\unA_{q}} = e^{i \int_{\S}\unB\unA^{g} }
\, ,
\ee
so that finally we have

\be
Z[\unA,\unB,A_{0},B_{0}]= e^{i \int_{\S \times [0]}\f F_{\unA} + i \int_{\S}
\unB\unA^{g}}\, .  \label{BFSIpart}
\ee

\subsubsection{Rederivation of solution on $\SS$}

In order to calculate the path integral on $\SS$ we put (\ref{BFSIpart})
into (\ref{BFpart}).
Putting the pieces together (and dropping the underlining), one arrives 
at the partition function
\be
Z_{\SS} = \int DA\, Dg\, DB\, D\f\, e^{iS}   
\ee
where the action is
\be
S= \int_{\S} \f F_{A} + B (A^{g} - A)\, . \label{BF2action}
\ee

Notice that we have 
exchanged the measure $DA_{0}\Det'(D_{0})\left.\right|_{0}$
for $Dg$. This is in fact correct as 
$\Det{}'(D_{0})\left.\right|_{0} = \Det((1-\Ad (g))/ \ad(A_{0}))$ 
which is the required Jacobian. Notice also that the change of variables
from $B_{0}$ to $\f$ produces exactly the right Jacobian to cancel the 
remaining $\Det'(D_{0})\left.\right|_{0}$.

The action still has a great
deal of symmetry. Conventional gauge invariance is there as well as
invariance under
\be
\d B = (\ed + \frac{1}{2}\ad (A + A^{g}))\Lambda \, , \; \;
\;  \d \f = g \Lambda g^{-1} - \Lambda \, . \label{newgt}
\ee

We can conjugate the group field $g$ into the torus of the group $t \in T $.
Once we have done this, as usual, all the non-trivial torus bundles are
liberated. We must sum over all of the possible first Chern classes
associated with these. As a next step of gauge fixing we also wish to
fix $\f$ to lie in the Cartan subalgebra $\lt$, the Lie algebra of
$T $. This is done by making use of the symmetry (\ref{newgt}). It is
important to notice that this gauge is not achieved by conjugation but,
rather, by a shift
\be
\f^{\lk} \rightarrow \f^{\lk} + (1-\Ad (t))\Lambda^{\lk}
\ee
so that one is not `mixing' torus bundles. Also note that the $\lt$ part
of $\Lambda$ does not appear in this transformation rule. 

With $g$ in the torus,
$B^{\lk}$ appears in the action solely in the term $B^{\lk}(1-\Ad
(t))A^{\lk}$ so that on integrating it out we obtain
\be
\Det{}_{\lk}(1-\Ad (t))\left.\right|_{1}^{-1}\d(A^{\lk})\, .
\ee
The $B^{\lt}$ integral sets $t$ to be position independent (modulo
infinities from the harmonic modes, whose existence depends on whether we
allow $t^{-1}\ed t$ to be non-exact). Furthermore,
the integral over the torus component of the gauge field, including a
sum over all possible Chern classes, gives a delta function constraint
onto constant and integral $\f$. Thus from the gauge fixing for $g$, the
gauge fixing for $\f$ and the final part of the functional integral we pick
up exactly the right determinants to reproduce the result
(\ref{resultBFSS}). Note that if we had chosen the
gauge fixing $D_{0}B_{0} = 0$ instead of $\partial_{0}B_{0} = 0$, the 
ghost determinant
would now be $\Det(D_{0})\left.\right|_{0}^{2}$. Solving the gauge constraint
would force $\f$ to lie in the torus spanned by $A_{0}$ (when coupled with
periodicity) and changing variables to eliminate this constraint would
produce the determinants 
$(\Det(\ad (A_{0}))\left.\right|_{0} \Det(D_{0})\left.\right|_{0})^{-1}$
which are just 
sufficient to leave an overall $\Det'(D_{0})\left.\right|_{0}$,
the determinant we obtain from (\ref{BF2action}) after gauge fixing $\f$ to
lie in the torus.

\subsection{$U(1|1)$ model}

Rozansky and Saleur introduced a cohomological field theory in
\cite{RozSal} which
they related to the Alexander polynomial. In this way they were able to
give a field theoretic proof of the relationship between the Alexander
polynomial and the Ray-Singer Torsion. The model they consider is a cousin
to the topological field theory used to describe the Casson invariant.
Indeed it is a type of $U(1)$ version of the Casson model (also known as
three-dimensional super $BF$ theory \cite{physrep}), the
conventional $U(1)$ Casson model being trivial. In our previous
examples the bundles were, from the outset, trivial. In
principle we should now specify which $U(1)$ bundle we are talking
about. However, the path integral has delta function support on flat
connections and so we may as well fix our attention on the trivial
$U(1)$ bundle. 

The action is
\be
\int_{M}\,B\ed A + \pb (d + A) \p \, , \label{rsact}
\ee
and has an $N=2$ topological supersymmetry as well as conventional gauge
invariance,
\bea
& & \d A = d\o \, , \; \; \; \; \; \; \; \d B = \ed\rho - \pb \sigma +
\bar{\sigma}\p  \, , \nonumber \\
& & \d \p = (\ed + A) \sigma \, , \; \; \; \d \pb = (\ed -A) \bar{\sigma} 
\, . \label{trrs}
\eea

\noi Rozansky and Saleur employ conformal field theory techniques to compute
the partition function of this theory. We will reproduce their results by a
direct path integral calculation.

Our choice of gauge on a three manifold $\SS$ is
\be
\dot{A_{0}}=0 \, , \; \; \; \dot{B_{0}}=0 \, \; \; \;  (\partial_{0}
-A_{0})\p_{0} =0 \, , \; \; \; (\partial_{0} + A_{0}) \pb_{0} =0 \, .
\label{gfrs}
\ee
The $B$ zero modes do not enter into the theory at all, and correspond to
symmetries which may be gauge fixed to zero. This is understood to have been
done. Here we have no problems with zero modes as the shift symmetry 
$\d B = \ed \rho$ is manifest in the original action. 

In order to implement these gauge choices one needs to append to the
action (\ref{rsact})
\be
\int_{\SS}\, E \partial_{0} A_{0}+ \bar{E} \partial_{0} B_{0} + \eb
(\partial_{0} -A_{0})\p_{0} + \e (\partial_{0} + A_{0}) \pb_{0} \, , 
\ee
as well as the following Faddeev-Popov ghost terms
\bea
&& \int_{\SS}\,\bar{\o}\partial_{0}\o + \bar{\rho}(\partial_{0} +\bar{\sigma}
\p_{0}  -
\pb_{0} \sigma) + \fb (\partial_{0} -A_{0})(\partial_{0}+A_{0})\sigma 
\nonumber\\ && + 
\f (\partial_{0} +A_{0})(\partial_{0}-A_{0})\bar{\sigma} \, .
\eea

Integrating over the $B_{i}(t)$ field we obtain a delta function
constraint
\be
\d(\partial_{0}A_{i}(t) - \partial_{i}A_{0}) \, ,
\ee
which again, together with periodicity in the $S^{1}$ variables, tells
us that $A_{i}$ is time independent and that $A_{0}$ is constant (and will
henceforth be renamed $a_{0}$ to make this explicit). As
usual, we have the freedom to fix $a_{0}$ to lie on the circle and we do
so. Now as $a_{0}$ is constant, and as long as it is not zero, one
deduces from the gauge fixing conditions (\ref{gfrs}) that
\be
\p_{0}= \pb_{0}=0 \, . \label{eq:psitozero}
\ee
The path integral thus reduces to a product of the determinants
\be
\frac{{\Det{}'}^{2}(\partial_{0})\left.\right|_{0}}
     {{\Det{}}'    (\partial_{0})\left.\right|_{1}} 
\, . \, 
\frac{{\Det}(\partial_{0} + a_{0})\left.\right|_{1}}
     {{\Det}(\partial_{0} + a_{0})(\partial_{0}-a_{0})\left.\right|_{0}}
\label{rsdets} 
\ee
times 
\be
\int \,DB_{0}DA \exp{i \left( \int_{\S} \,B_{0}dA \right)} \, .
\ee

The ratio of determinants (\ref{rsdets}) is essentially
\be
\int_{0+\epsilon}^{2\pi-\epsilon}da_{0}\,(2\sin{a_{0}/2})^{-\chi(\S)} \, . 
\label{rsint} 
\ee
For $g \geq 1$ one may dispense with the regularisation and in this case
we obtain
\be
V_{M}({\cal M}) = 2^{2g-2} \left( \begin{array}{c}
                            2g-2 \\ g-1
                           \end{array}\right) \, . \label{eq:U(1,1)result} 
\ee
where $V_{M}({\cal M})$ is the volume of the moduli space of flat
connections with the Ray-Singer torsion as the measure. In fact we can
absorb overall factors $a.b^{g-1}$ into the normalization \cite{btver}, and
it is the combinatorial factor that is interesting. 
This result is in agreement with Rozansky and Saleur \cite{RozSal}. 
One can also derive an equivalent two-dimensional description of the $U(1|1)$
model along the lines of the calculation leading to (\ref{BF2action}). 

In passing, we want to point out that the combinatorial factor
${2g-2 \choose g-1}$ can be interpreted as the Euler character of
the $(g-1)$'th symmetric power of $\S$. This result agrees with a 
calculation of the Seiberg-Witten-Casson invariant \cite{swc}
when the spin-c bundle is the canonical line bundle of $\S$. 

\section{The mapping torus $\maptor$}
\label{sec-maptor}

Thus far our attention has been on manifolds of the type $\SS$. We may
generalize somewhat and nevertheless keep solvability by passing to 
mapping tori.

Let $\beta: \S \maps \S$ be a diffeomorphism of $\S$. Associated with 
$\beta$ and $\S$ is a three-manifold $\maptor$ known as the mapping torus of
$\beta$. 
It is obtained from the manifold $\SI$ by making 
the following identification: $(\beta x,0) \sim (x, 1)$. This is clearly 
a fibration over $S^{1}$ with the obvious projection. Fields on $\maptor$ may
be regarded as fields on $\SI$ subject to the twisted periodicity condition
\be
\phi(t + 1) = \beta^{*}\phi(t)\, .
\ee
We will need to know one basic fact about the cohomology of $\maptor$. First
of all, $\beta$ induces an action on the cohomology of $\S$ (denoted
$\beta^{*}$ as it is essentially a pullback). The degree one cohomology is
then generated by the pullback of the generator of $H^{1}(S^{1})$ and by the
fixed points of the action of $\beta^{*}$ on $H^{1}(\S,\RR)$. If there are
no such fixed points, in other words when $(\beta^{*} - 1)$ is invertible,
then $H^{1}(\maptor,\RR)$ is one-dimensional, and this is the case we will
consider henceforth. In genus one this will be the case whenever the induced
$SL(2,\ZZ)$ matrix $U_{\beta}$ satisfies $\Tr(U_{\beta}) \neq 2$. It should
be noted however that the reduction to two dimensions does not rely on this
assumption,and the actions we obtain are therefore valid for general $\beta$. 

\subsection{$G/T$ model}

The model we want to consider is very closely related to the $U(1|1)$ model
above but with a slightly different fermionic sector. The matter
fields (which in fact we will take to be commuting or anti-commuting) take
values in $\frak{k}$ and transform under the adjoint action of $T$.  
For simplicity we will take the fields as having values in 
the $\lk$ directions in $SU(2)$ where the gauge group is $U(1)$ although
everything goes through in exactly the same way in the more general case. 
The reason for the choice of the adjoint action is that the determinant is
always unity and thus there is no problem in defining the phase of the path
integral, a subtlety that also arises in Chern-Simons theory, reflecting 
the problem of choosing a framing. 

The theory is then given by
\be
S(A,B,\p,\pb) = 
       \int_{M} \,B\ed A + \pb(\ed + \ad(A))\p \label{eq:GTaction}\,
	   .
\ee

\noi This has the following infinitesimal symmetries depending on the
statistics of $\p$ and $\pb$:
\be
\begin{array}{ll}
\delta A = \ed\omega & \delta\p = -\ad(\omega)\p + \ed_{A}\sigma \\
\delta B = \ed\rho - (-1)^{\p}(\ad(\bar{\sigma})\p + \ad(\sigma)\pb) &
		\delta\pb = -\ad(\omega)\pb + \ed_{A}\bar{\sigma}
\end{array}\, .
\ee

The partition function for (\ref{eq:GTaction}) is given by
\be
Z = \int_{T^{b_{1}}} T^{RS}_{M}(g)^{\pm 1} \label{eq:GTpart}
\ee

\noi where $b_{1}$ denotes the first Betti number of $M$ and 
the parameter $g$ is given by the holonomy along
the non-trivial cycles of the manifold. We note that a straightforward
generalization of the argument used for the $U(1|1)$ model would enable us
to solve this model for $M = \SS$ but now we concentrate on the mapping
torus.

\subsubsection{Gauge fixing}

The first issue we have to deal with on $\maptor$ is that of gauge fixing.
We want to ``project out'' the two dimensional theory by gauge fixing away
the third dimension. This was achieved for $\SS$ by gauge fixing $A_{0}$ 
to the class of functions constant in time. The situation is more
complicated for $\maptor$ as we are not now dealing with a product manifold,
and a global definition of a ``time'' direction cannot be given. To make
things clearer we shall work on $\S \times \RR$ with functions that satisfy
the condition $\beta^{*}\phi (x,t) = \phi (x,t+1)$, or alternatively on the
interval with functions satisfying $\beta^{*}\phi (x,0) = \phi (x,1)$.
It is immediately obvious that we cannot simply gauge fix to the class of
time-constant functions, as they will not neccessarily satisfy this
condition. Instead we must look for a class of functions that do satisfy the
condition, but which is no ``larger'' than the time-constant class. One
possibility is to multiply a constant function by a time dependent part that
vanishes at the boundary of the interval (or at every integer if we are
thinking of the real line). We need to achieve this gauge via a gauge
transformation that is the identity at the boundary of the interval (we have
boundary conditions to preserve). In the present (abelian) case we find that
the requisite gauge parameter is
\be
\omega (x,t) = \int_{0}^{t} ds\, (f(s)\hat{A}_{0}(x) - A_{0}(x,s)).
\ee
The requirement that it vanish also at $t=1$ imposes the following
condition:
\be
\hat{A}_{0}(x) \int_{0}^{1}dt\, f(t) = \int_{0}^{1}dt\, A_{0}(x,t).
\ee
The simplest thing is to normalize $f$ to integrate to unity over the 
interval. Thus we can achieve our gauge choice, still parametrized by a
function on $\S$, using suitable gauge transformations. The gauge fixing
introduces two determinants of $D_{0}$. Given that the gauge group is
abelian we have no need of these to provide Jacobians for a change of
measure to the group. In fact they are cancelled, as in the case of $BF$
theory on $\SS$, by a field redefinition similar to that in equation
(\ref{redefB}).

\subsubsection{Reduction to 2 dimensions}

We proceed as we did for the reduction to two dimensions of $BF$ theory,
starting on $\SI$ specifying $\underline{\p}$ and $\unA$ at $t=0$ and 
$\underline{\pb}$ and $\unB$ at
$t=1$. Using the results about gauge fixing above 
we then find the following two-dimensional action (the derivation is not
completely trivial and will be explained in \cite{us}):
\be
S = \int_{\S}\,B_{0}\ed A + \pb(\beta^{*} - \Ad(g))\p +
    \p_{0}\ed_{\beta^{*}A}\pb + \pb_{0}\ed_{A}\p +
    B(\beta^{*}A - A - g^{-1}\ed g) \label{eq:GT2dimaction}
\ee

\noi where we omit the underlining on the now spatial one-forms $\unA$,
$\unB$, $\underline{\p}$ and $\underline{\pb}$. $g$ here is $e^{A_{0}}$.
While this action has an explicit dependence on $\beta \in \Diff{\S}$,
it can be shown \cite{us} that the theory depends only on the smooth manifold
$\maptor$ and is, in particular, invariant under isotopies of $\beta$. 
We note that there is a symmetry of the form $\d B = \ed \s$ with
compensating terms in $\d B_{0}$, but that this is independent of the fields
and will henceforth be ignored (assumed gauge fixed). 

This action has the following local symmetry:
\be
\begin{array}{lll}
\p \mapsto h^{-1}\p h & 
\pb \mapsto \beta^{*}h^{-1}\p \beta^{*}h &
g\mapsto h^{-1}g\beta^{*}h \\
\p_{0} \mapsto \beta^{*}h^{-1}\p_{0} \beta^{*}h & 
\pb_{0} \mapsto h^{-1}\pb_{0}h & A \mapsto A + h^{-1}\ed h 
\end{array}\, .
\ee

\noi The somewhat unusual action of the gauge group on the group valued field
$g$, $g\mapsto h^{-1}g\beta^{*}h$, will also appear in the analogous 
two-dimensional models for $BF$ and Chern-Simons theory to be discussed below.
To solve we make the following steps:
\begin{enumerate}
\item We gauge fix the symmetry using $\ed\ast A = 0$. This condition
together with the flatness condition imposed by the $B_{0}$ integral tells
us that $A$ is harmonic. Note that this statement has not required the
integral over the $B_{0}$ zero mode. 

\item The $B$ integral is telling us that 
$(\beta^{*} - 1)A = g^{-1}\ed g$. Because of the gauge fixing we know that
$A$ is an harmonic form, call it $\omega$. The exterior derivative commutes
with the pullback and thus, by the Hodge decomposition we know that 
$\beta^{*}A$ can be written as
\be
\beta^{*}A = \gamma + \ed \alpha
\ee

\noi where $\gamma$ is harmonic (in the original metric) and $\alpha$ 
will of course depend on $\omega$. The right hand side
$g^{-1}\ed g$ (as $\int_{0}^{1}A_{0}$ is globally well-defined) is exact 
and so from orthogonality we are led to the two
equations $\gamma = \omega$ and $\ed\alpha = g^{-1}\ed g$. Now 
making the assumption that $b_{1}=1$, so that $(\beta^{*} - 1)$ on
the harmonic forms is invertible, we can conclude that  $\omega = 0$. 
Thus $A$ and its pullback are zero and hence $g^{-1}\ed g$ is zero as well. 

\item There is an extra symmetry to be gauge fixed viz:
\be
\begin{array}{ll}
\delta\p = \ed_{A}\rho & \delta B_{0} = [\pb_{0},\rho ] \\
\delta\p_{0} = - (-1)^{\p}(\beta^{*} - \Ad(g))\rho &
\d B = - (-1)^{\p} [\Ad(g)\rho,\pb]
\end{array}  \label{eq:GTsym2}
\ee

\noi that depends on the statistics of $\p$, and an analogous one for 
$\pb$. These we fix by the
conditions $\ed_{A}\ast\p = 0$ and $\ed_{A}\ast\pb = 0$, although
we already know that $A = 0$ in fact. Then the $\p_{0}$ and
$\pb_{0}$ integrals tell us that $\p$ and $\pb$ are harmonic
also. 

\item During the gauge fixing of $\p$ and $\pb$ we have not used
the $\rho$ zero mode as $\delta\p = \ed\rho$, so there are still constant
$\rho$ gauge transformations which leave the gauge fixed action invariant.
Notice also that $\p_{0}$ constant modes do not appear in the action 
(\ref{eq:GT2dimaction}). This happy situation means that providing 
$\Det (1- \Ad(g)) \neq 0$ we can use the constant $\rho$ symmetry by 
(\ref{eq:GTsym2}) to gauge fix the $\p_{0}$ constant modes to zero and we do
so. A similar story holds for $\pb_{0}$.

\item From the gauge fixing of $\p_{0}$ and $\pb_{0}$ we pick up ghost
determinants
\be 
\det(1 - \Ad(g))_{H^{0}(\S)}^{\mp 2}.
\ee

\item Thus all except the $g$ integral have been performed and we are left
with the following expression for the partition function of the theory on
$\maptor$: 
\be
Z[\beta] = \int \,dg \left\{ \det (\beta^{*} - \Ad(g))_{H^{1}(\S)}\cdot
                          \det (1 - \Ad(g))_{H^{0}(\S)}^{-2}
						  \right\}^{\pm 1}
\ee

\noi where the sign of the exponent depends on the statistics of
$\p$. Comparing this equation to (\ref{eq:GTpart}) allows us to claim that
on a mapping torus, the Ray-Singer torsion is
\be
T_{\maptor}(g) = \det (\beta^{*} - \Ad(g))_{H^{1}(\S)}\cdot
                          \det (1 - \Ad(g))_{H^{0}(\S)}^{-2}
\ee

\noi and this agrees with the result of Fried \cite{fried}.
						  
\end{enumerate}

\section{Conclusions and future work}

We briefly discuss the theories one obtains on reducing $BF$ and
Chern-Simons theories on the mapping torus to equivalent two dimensional
theories.

Starting from the action for $BF$ theory we can follow a similar procedure
to that in the $\SS$ case to reduce the theory on $\maptor$ to two
dimensions. The result is that the two dimensional action becomes
\be
S = \int_{\S}\f F_{A} + B(A^{g} - \beta^{*}A) \label{BF2twistaction}
\ee
with gauge symmetries
\be
\begin{array}{ll}
A \mapsto h^{-1}Ah + h^{-1}\ed h & g \mapsto h^{-1}g\beta^{*}h  \\
\f \mapsto h^{-1}\f h  & B \mapsto \beta^{*}h^{-1}B\beta^{*}h
\end{array}\, .\label{ggfsymm}
\ee
and extra symmetries
\be
\begin{array}{ll}
\d \phi = g\Lambda g^{-1} - (\beta^{*})^{-1}\Lambda &
\d B = [\ed + \frac{1}{2} \ad(A^{g} + \beta^{*}A)]\Lambda
\end{array}
\ee
In order to evaluate the partition function of (\ref{BF2twistaction}) we
cannot employ abelianization directly. Instead one makes use of a
localization argument to proceed.
For Chern-Simons theory on a mapping torus one can also obtain
an equivalent two-dimensional theory. One finds a $\beta$-twisted $G/G$ model,
i.e.\ an anomaly-free  $G_{L} \times G_{R}$ gauged $WZW$ model
(see \cite{witt5}) with $A_{R} = \beta^{*}A_{L}$
and local gauge invariance given by the first line of (\ref{ggfsymm}).
These two-dimensional theories derived from $BF$ theory and Chern-Simons
theory again explicitly involve a diffeomorphism $\beta\in\Diff{\S}$, and
again only depend on the conjugacy class of the isotopy class of this
diffeomorphism.

We have seen that for various admittedly simple three manifolds a direct
gauge-theoretic evaluation of partition functions of topological field 
theories is possible. The same techniques allow one to calculate 
observables in these theories as well. 

For the theories considered here we have at no time explicitly needed to 
introduce a metric on $\S$. However for Chern-Simons theory on $\SI$ the
boundary data {\em was} specified relative to a metric \cite{btver}. At
the end one proves that nothing depends on this choice. One can view the $BF$
theories as special cases of Chern-Simons theory so that for example when
$\lambda = 0$ one would find be the $IG/IG$ gauged $WZW$ model as
the two-dimensional equivalent. 
The relationship between such two-dimensional actions 
and the manifestly metric-independent actions obtained here is
akin to the relationship between the $G/G$ model and the gauged $WZ$ 
term described in \cite{witt4} and \cite{btequivloc}. The details of all
these observations, as well as a comparison with the work on Chern-Simons
theory on a mapping torus in \cite{jeffrey}, will be described in \cite{us}.

\end{document}